\documentclass[twocolumn,showpacs,preprintnumbers]{revtex4}
\usepackage{graphicx}
\usepackage{dcolumn}
\usepackage{bm}

\input{tcilatex}

\begin{document}

\title{Effects of local oxygen distortions on electronic structures of Na$%
_{x}$CoO$_{2}$}
\author{Jun Ni$^{1}$ and Guang-Ming Zhang$^{2}$}
\affiliation{$^1$Department of Physics, Tsinghua University, Beijing 100084, China;\\
$^{2}$Center for Advanced Study, Tsinghua University, Beijing 100084, China}
\date{\today }

\begin{abstract}
By using pseudopotential method with local spin density functional
approximation, the electronic band structures of Na$_{x}$CoO$_{2}$ are
calculated for $x=0.25$, $0.5$, $0.75$, and $x=1$ in the presence of the
structure relaxations. As increasing Na content, the hybridization between
cobalt and oxygen orbitals is decreased, and a phase transition is predicted
from a wide-band ferromagnetic to a narrow band paramagnetic metals. The
itinerant ferromagnetism is strongly suppressed by the local distortions of
the oxygens around the cobalts. Moreover, straining the CoO$_{2}$ layers
corresponding to the hydrated superconductor Na$_{0.35}$CoO$_{2}\cdot $1.3H$%
_{2}$O strongly enhances both the hybridization and ferromagnetism.
\end{abstract}

\maketitle

Since the discovery of high critical temperature superconductivity in
layered copper oxide compounds, the search for layered superconductors
without copper atoms has been regarded as an important route to understand
the unknown mechanism behind the highest superconducting transition
temperatures. Recently, Takada and coworkers \cite{takada} have discovered a
novel superconductor Na$_{0.35}$CoO$_{2}\cdot $1.3H$_{2}$O with T$_{c}=5$K,
which is synthesized when a sodium cobalt oxide Na$_{0.75}$CoO$_{2}$ is
hydrated to reduce its sodium content and intercalated with water molecules
to further separate the CoO$_{2}$ layers. Based on the analogy of its
layered transition metal oxide structure, it has been speculated that the
displayed superconductivity may be closely related to that of the high T$%
_{c} $ cuprates \cite{takada,sakurai}. Along this line, several theoretical
analysis \cite{baskara,shastry,wang} have been carried out in terms of
resonant valence bond states on a strongly correlated two-dimensional t-J
model on triangular lattices.

However, unlike the parent compounds of high T$_{c}$ cuprates as an
antiferromagnetic Mott insulator, the parent compounds Na$_{x}$CoO$_{2}$ ($%
x<1$) have been known for their anomalous large thermoelectric power and low
electronic resistivity \cite{terasaki}. Na$_{x}$CoO$_{2}$ consists of a
two-dimensional triangular lattice of cobalt ions formed by a network of
edge-sharing CoO$_{6}$ octahedra, separated by layers of sodium ions. Band
structure calculations using a general potential linearized augmented plane
wave method \cite{singh2000} for Na$_{0.5}$CoO$_{2}$ indicates that it is a
ferromagnetic half metal with little hybridization between the oxygen and
cobalt orbitals, while no distinct magnetic ordering has been observed
experimentally \cite{terasaki,ando}. But for the samples of Na$_{0.75}$CoO$%
_{2}$, a magnetic phase transition associated with the antiferromagnetic
interactions has been confirmed below $T=22$K, being accompanied by a
specific heat jump, a small magnetization, and a kink in the temperature
dependence of the resistivity \cite{motohashi,sugiyama}.

Moreover, it has been reported that the crystal structure of the
superconducting Na$_{0.35}$CoO$_{2}\cdot $1.3H$_{2}$O compound only differs
from the parent compound Na$_{0.75}$CoO$_{2}$ by the intercalation of water
molecules to enlarge the CoO$_{2}$ layers and to reduce the sodium charges %
\cite{takada,cava}. It is thus of great interest to scrutinize the
electronic band structures of the parent compound Na$_{x}$CoO$_{2}$ for
different sodium contents and to study the electronic structure with a large
CoO$_{2}$ layers as the same as that of the hydrated compound.

In this paper, by using pseudopotential method with local spin density
functional approximation \cite{kohn-sham,barth,perdew-wang}, we present the
first principles electronic structure calculations on Na$_{x}$CoO$_{2}$ for
different Na content. The electronic structure is found to be highly two
dimensional, giving rise to a good metal for $x=0.25$, $0.5$, $0.75$ and a
semiconductor for $x=1$. In particular, at $x=0.25$ it is an itinerant
ferromagnetic metal with large hybridization between cobalt and oxygen
orbitals. As increasing the Na content, both the ferromagnetism and
hybridization are decreasing. When $x=0.5$, only tiny ferromagnetism is
left, and the hybridization vanishes. A phase transition is thus yielded
from the wide band ferromagnetic metal to a narrow band paramagnetic metal
by changing the Na content. The ferromagnetism is totally absent and a
narrow conduction band appears below the Fermi energy when $x=0.75$. We find
that such a strong suppression of the ferromagnetism is mainly originated
from the local distortions of the oxygen ions around the cobalt atoms.
However, the present treatments for the narrow band paramagnetic state may
underestimate the tendency of the material towards local moment formation
and possible magnetic ordering. In addition, straining the separation of the
CoO$_{2}$ layers hypothetically corresponding to the hydrated superconductor
Na$_{0.35}$CoO$_{2}\cdot $1.3H$_{2}$O, we find that both the hybridization
and ferromagnetism are enhanced, and the corresponding superconductivity
might be associated with a triplet (p-wave) pairing state in the presence of
strongly ferromagnetic fluctuations \cite{tanaka,singh2003}, analogous to
those in UGe$_{2}$, URhGe, and ZrZn$_{2}$ (Ref.\cite{saxena,aoki,pfleiderer}%
).

The crystal structure used in the present numerical calculations is based on
the neutron diffraction data reported by Balsys and Davis \cite{balsys}.
Neutron data for Na$_{0.74}$CoO$_{2}$ shows that it belongs to the hexagonal
P6$_{3}$/mmc space group with lattice parameters $a=2.840$\AA\ and $c=10.811$%
\AA . There are two CoO$_{2}$ sheets per unit cell. The triangles formed by
the oxygen ions in the oxygen layers have two directions. We use A and B to
represent these two oxygen layers. The oxygen packing in the structure is
AABB. The experimental results indicate that the oxygen atoms are at the
sites ($1/3,2/3,\pm 0.088$). The Na ions are intercalated with a trigonal
prismatic site between the octahedral sheets of cobalt oxide. There are two
available sites for the sodium ions: $(2/3,1/3,1/4)$ and $(0,0,1/4)$. The
sodium ions occupy these two sites unequally due to the influence of the Co
atoms above and below in the second layer. The probability of finding sodium
ions at the site $(2/3,1/3,1/4)$ being approximately twice that on the other
site $(0,0,1/4)$ \cite{balsys}.

In order to take into account the different stoichiometry $x=0.25$, $0.5$,
and $0.75$, an enlarged unit cell $(2\times 2\times 1)$ of the crystal
structure is introduced. The electronic Kohm-Sham orbitals for the momentum $%
\mathbf{k}$-points in the Brillouin zone are expanded in plane waves with a
cutoff of 24 Ryd. The PWSCF code \cite{baroni} is used in the calculations
with the ultrasoft pseudopotential \cite{vanderbilt} and Perdew-Wang
exchange correlation potential \cite{perdew-wang}. Using this method, we can
properly treat the atomic relaxations and consider their relations with the
electronic structures. We further assume the sodium ion setting on the site $%
(2/3,1/3,1/4)$. In fact, in our framework the position of sodium ions have
little effect on the electronic structures, because the role played by
sodium ions mainly donates charges to the CoO$_{2}$ layers. More
importantly, we have performed the structural relaxation to determine the
optimized atomic positions. In particular, it is found that there are
additional local distortions of the oxygen ions. While the earlier first
principle calculations \cite{singh2000,singh2003} did not consider such
local distortions.

During the calculations, we have noticed that the local distortions of the
oxygen ions strongly depend on the sodium content. As increasing the Na
content, the oxygen distortions are large, and the itinerant ferromagnetism
of Na$_{x}$CoO$_{2}$ is reduced significantly. To describe these local
distortions, we define that $z_{i}$ denotes the heights of the oxygen ions
of each layer in their z-direction to the Co layer in the unit of lattice
spacing $c$. All the oxygen layers (A and B) have the same magnitudes of the
local distortions. The relaxed positions of the oxygen ions have been
evaluated with respect to the sodium content and given in Table 1. We would
like to point out that the case with sodium ions on the site ($0,0,0$) has
similar relaxations for the oxygen ions.

{Table I. The oxygen positions and spin magnetizations per cobalt atom for
the unrelaxed (}$m_{u}${) and relaxed structures (}$m_{r}${) with respect to
the sodium content (x). }

\begin{center}
\begin{tabular}{c|c|c|c}
\hline\hline
x & $z_{i}(c)$ (i=1,2,3,4) & $m_{u}(\mu _{B})$ & $m_{r}(\mu _{B})$ \\ \hline
{x=0.25} & {z$_{1}$=0.0750; z$_{2}$,z$_{3}$,z$_{4}$=0.0787} & {0.775} & {%
0.340} \\ \hline
{x=0.5} & {z$_{1}$,z$_{2}$=0.0826; z$_{3}$,z$_{4}$=0.0773} & {0.534} & {0.018%
} \\ \hline
{x=0.75} & {z$_{1}$=0.0861; z$_{2}$,z$_{3}$,z$_{4}$=0.0824} & {0.263} & {0}
\\ \hline
{x=1} & {z$_{i}$=0.0921} & {0} & {0} \\ \hline\hline
\end{tabular}
\end{center}

The electronic band structures are displayed in Fig.1 and Fig.2 for $x=0.25$
and $x=0.75$, respectively. The energy scale is relative to the Fermi
energy, and the density of states is in atomic unit (Ryd)$^{-1}$. In Fig.1,
the left column corresponds to the spin minority electronic structure, and
the right column to the spin majority one. Both the spin minority and
majority electronic structures are the same for Na$_{0.75}$CoO$_{2}$. The
spin density of states (DOS) is calculated\ and delineated in Fig.3 for the
Na content $x=0.25$, $0.5$, $0.75$, and $1$. It is seen that a ferromagnetic
metal with the wide hybridized bands is obtained for $x=0.25$. For $x=0.5$,
there is a very weak ferromagnetism and little hybridization between the
cobalt and oxygen orbitals. Moreover, a paramagnetic metal with a narrow
band below the Fermi energy is found for $x=0.75$. By adding electron
charges to the CoO$_{2}$ layers, the Fermi energy approaches to the top edge
of the conduction band, and the difference between the spin minority and
spin majority is decreasing quickly. For the case of $x=1$, NaCoO$_{2}$
becomes a semiconductor, consistent with the earlier experimental result.
However, in the earlier density functional calculations \textit{without} the
local distortions of oxygen ions \cite{singh2000,singh2003}, a ferromagnetic
half metallic state is obtained for all the sodium contents in $x=0.3$, $0.5$%
, and $0.7$. In fact, we have also reproduced these results within the
present calculation scheme.

\begin{figure}[tbp]
\begin{center}
\includegraphics[width=3.0in]{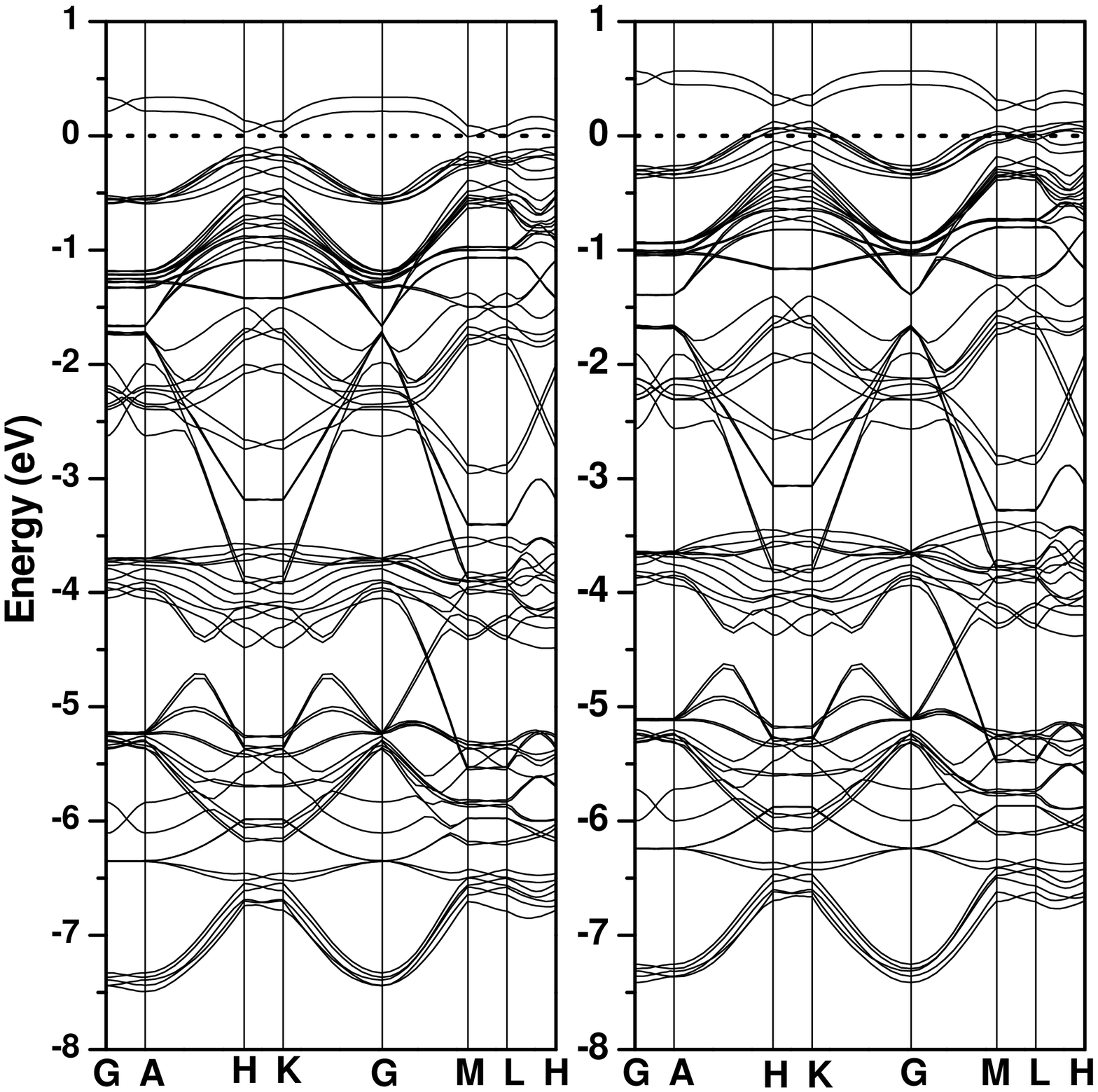}
\end{center}
\caption{The electronic band structure for Na$_{0.25}$CoO$_{2}$. The Fermi
energy $E_F$ has been set at zero and denoted by the dotted lines. The left
column corresponds to the spin-up electrons and the right column to the
spin-down electrons.}
\end{figure}

\begin{figure}[tbp]
\begin{center}
\includegraphics[width=2.7in]{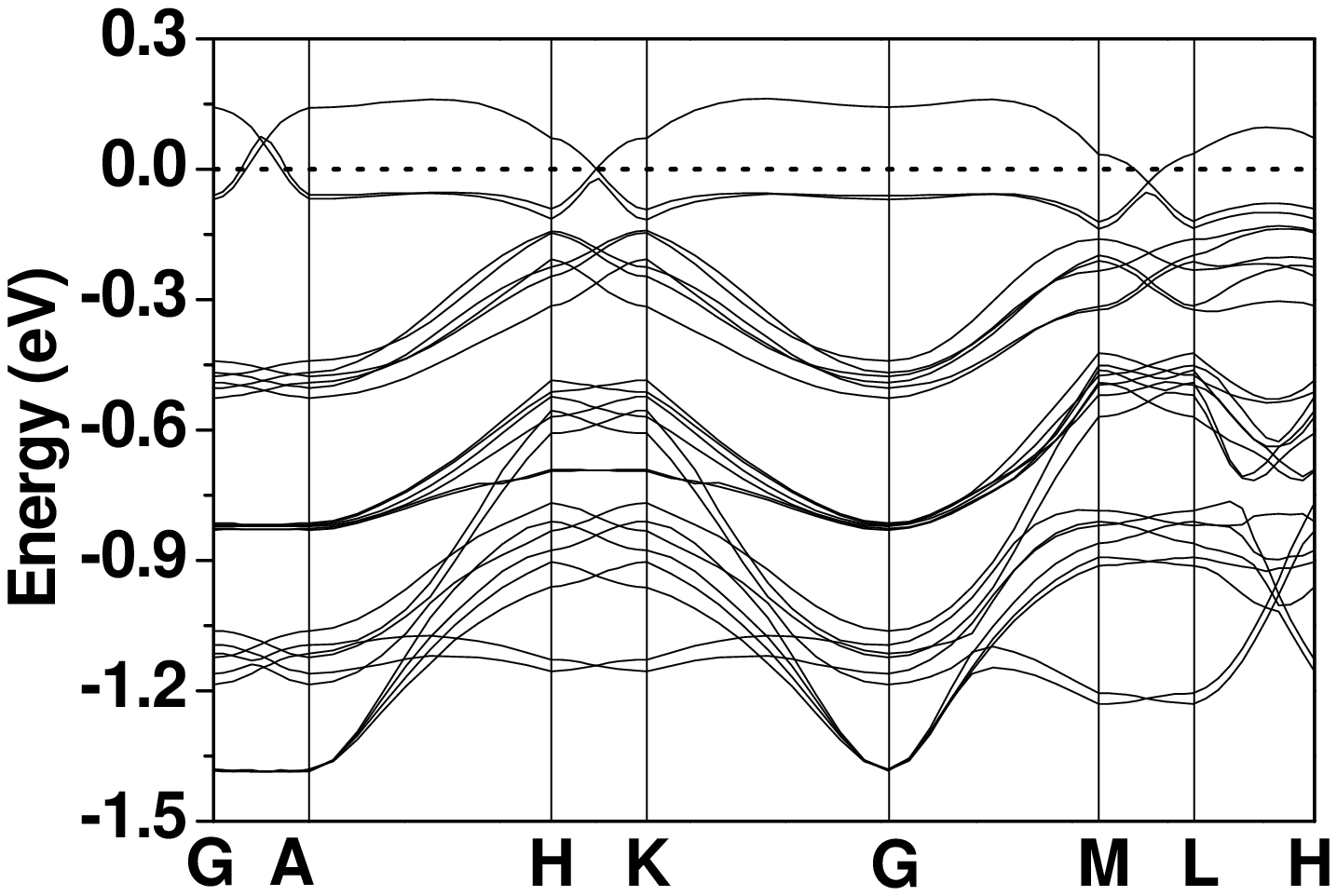}
\end{center}
\caption{The electronic band structure for Na$_{0.75}$CoO$_{2}$. There are
no differences between the spin-up and spin-down electrons. The Fermi energy 
$E_F$ has been set at zero and denoted by the dotted lines.}
\end{figure}

Moreover, the local spin density functional calculations can also
self-consistently determine the spin magnetization of the ferromagnetically
ordering state for low Na content ($x\leq 0.5$). In Table 1, the
corresponding spin magnetizations per Co atom are given for both unrelaxed
and relaxed structures. For $x=0.5$, our result with the \textit{unrelaxed}
oxygen position is $0.534\mu _{B}$ per cobalt atom. However, in the \textit{%
relaxed} structure of material, it is the local oxygen distortions that
reduce the spin magnetization to $0.018\mu _{B}$ per cobalt ion. This result
is consistent with the experimental observation \cite{terasaki} and the
nominal valence of $+3.5$ of each cobalt ion in Na$_{0.5}$CoO$_{2}$.
Therefore, it is expected that the material Na$_{0.5}$CoO$_{2}$ is in the
vicinity of the critical sodium content separating the weakly coupling wide
band ferromagnetic metal from the strongly coupling narrow band paramagnetic
metal. Particularly, the Coulomb interaction in Na$_{0.75}$CoO$_{2}$ may be
larger than the conduction electron bandwidth, and the present spin density
functional approximation may not be adequate to considering the possible
antiferromagnetic instability.

\begin{figure}[tbp]
\begin{center}
\includegraphics[width=3.0in]{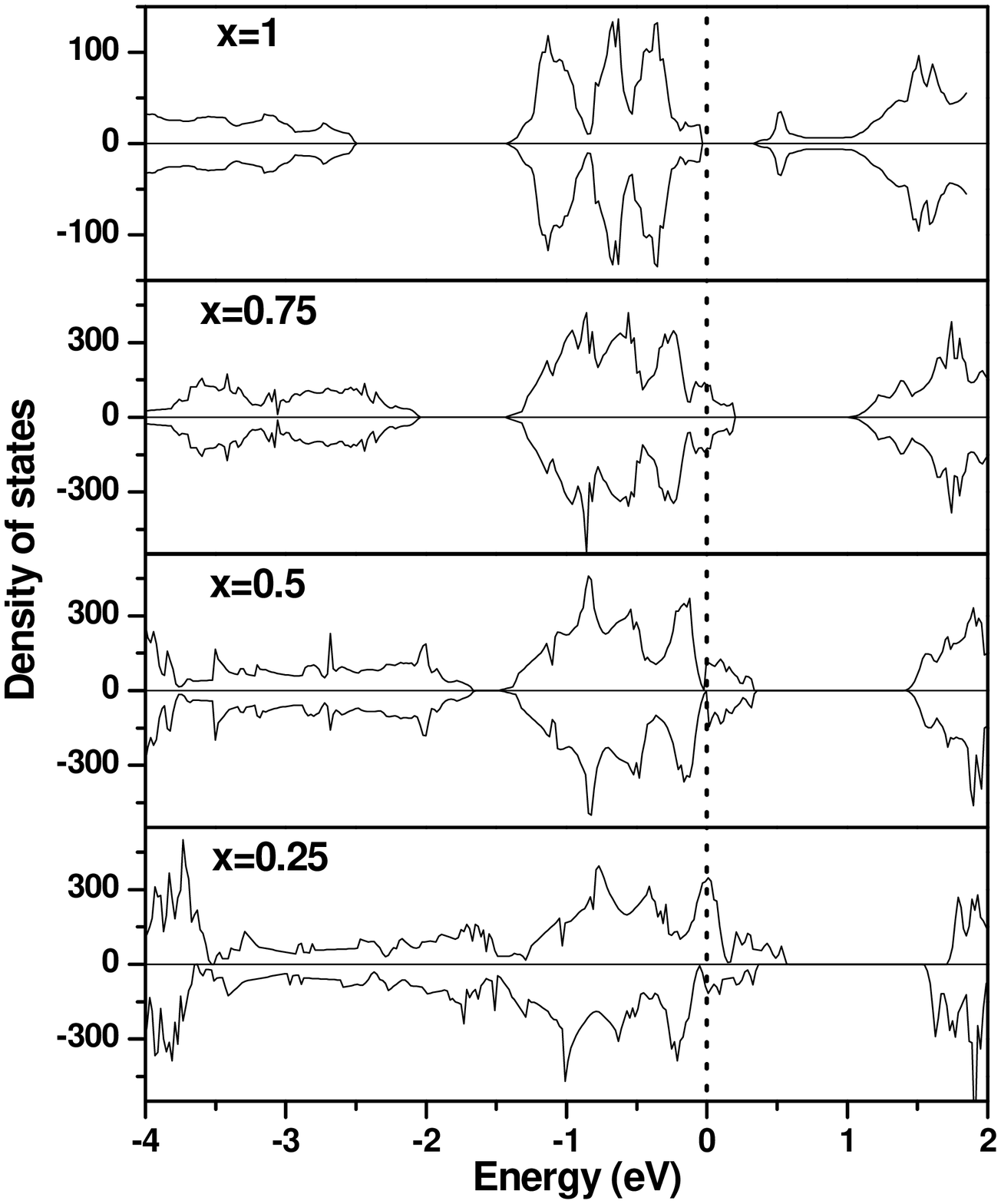}
\end{center}
\caption{The density of states of Na$_{x}$CoO$_{2}$ for the different sodium
content with relaxed structure. The density of states is in atomic unit (Ryd)%
$^{-1}$. The up-spin DOS is shown above the horizontal line and the
down-spin below.}
\end{figure}

In order to exhibit the anisotropy of the charge distribution, we also
calculate the local density of states (LDOS) at Fermi energy defined by $%
n(E_{F},\mathbf{r})=\sum_{i}|\psi _{i}(\mathbf{r)}|^{2}\delta (E_{F}-E_{i})$%
, which can provide some insights into the electron bonding around the
individual atoms. The LDOS at the Fermi energy contour maps for the (001)
surface are given in Fig.4. For both Na$_{0.75}$CoO$_{2}$ and Na$_{0.25}$CoO$%
_{2}$, Co ions have a spherically symmetric density distribution around each
site on the Co layer (Fig.4b and 4e), while their local density
distributions change from a down-triangular symmetry below Co layer (Fig. 4a
and 4d) to an up-triangular symmetry above Co layer (Fig.4c and 4f). The
down- and up-triangles in the lower sodium content have more sharper edges
than the higher sodium content, and the oxygen ions with a spherically
symmetric distribution have higher density in low sodium content as well.
There is a little charge density around Na ions, reflecting that the role of
Na atoms is mainly to donate charges to the CoO$_{2}$ layer. These results
clearly exhibit that a mixed metallic and covalent bonding are associated
with the hybridization between the cobalt and oxygen orbitals, and that the
compounds are clearly quasi-two dimensional systems.

\begin{figure}[tbp]
\begin{center}
\bigskip
\end{center}
\caption{The spatial distributions of the local density of states at the
Fermi energy for the (0,0,z) surface in unit of c. For Na$_{0.75}$CoO$_{2}$,
(a) $z=-0.0370$, (b) $z=0.0$, (c) $z=0.0462$. For Na$_{0.25}$CoO$_{2}$, (d) $%
z=-0.0370$, (e) $z=0.0$, (f) $z=0.0462$.}
\end{figure}

Finally, to understand the hydration effects of the compounds Na$_{x}$CoO$%
_{2}$ ($x<1$) on the electronic band structures and the corresponding
superconducting mechanism, we strain the separation of the CoO$_{2}$ layers
for Na$_{0.25}$CoO$_{2}$ and Na$_{0.75}$CoO$_{2}$ without taking into
account the water molecules. The corresponding DOS with $c=3.47$\AA\ are
displayed in Fig.5. It is clearly seen that the hybridization between the
cobalt and oxygen orbitals are considerably increased and the ferromagnetism
prevails even in the high sodium content. Na$_{0.25}$CoO$_{2}$, for example,
the corresponding spin magnetization is increased to $0.785\mu _{B}$, nearly
two times of the corresponding value with normal separation of CoO$_{2}$
layers. Thus, an effective one-band model Hamiltonian may exist to describe
the obtained electronic structure of this material. Actually, the
singularity in the DOS seen in Fig.5 can induce strong quantum ferromagnetic
fluctuations, from which an effective attraction among itinerant electrons
is generated as the same as the conventional low-temperature superconductors
driven by the electron-phonon interaction \cite{machida,belitz,roussev,santi}%
. The possible mechanism of superconductivity in sodium cobalt oxides may
thus originate from the formation of a triplet (p-wave) electron pairing
state. The similar mechanisms have been proposed for the materials UGe$_{2}$%
, URhGe, and ZrZn$_{2}$ (Ref.\cite{saxena,aoki,pfleiderer}).

\begin{figure}[tbp]
\begin{center}
\includegraphics[width=3.0in]{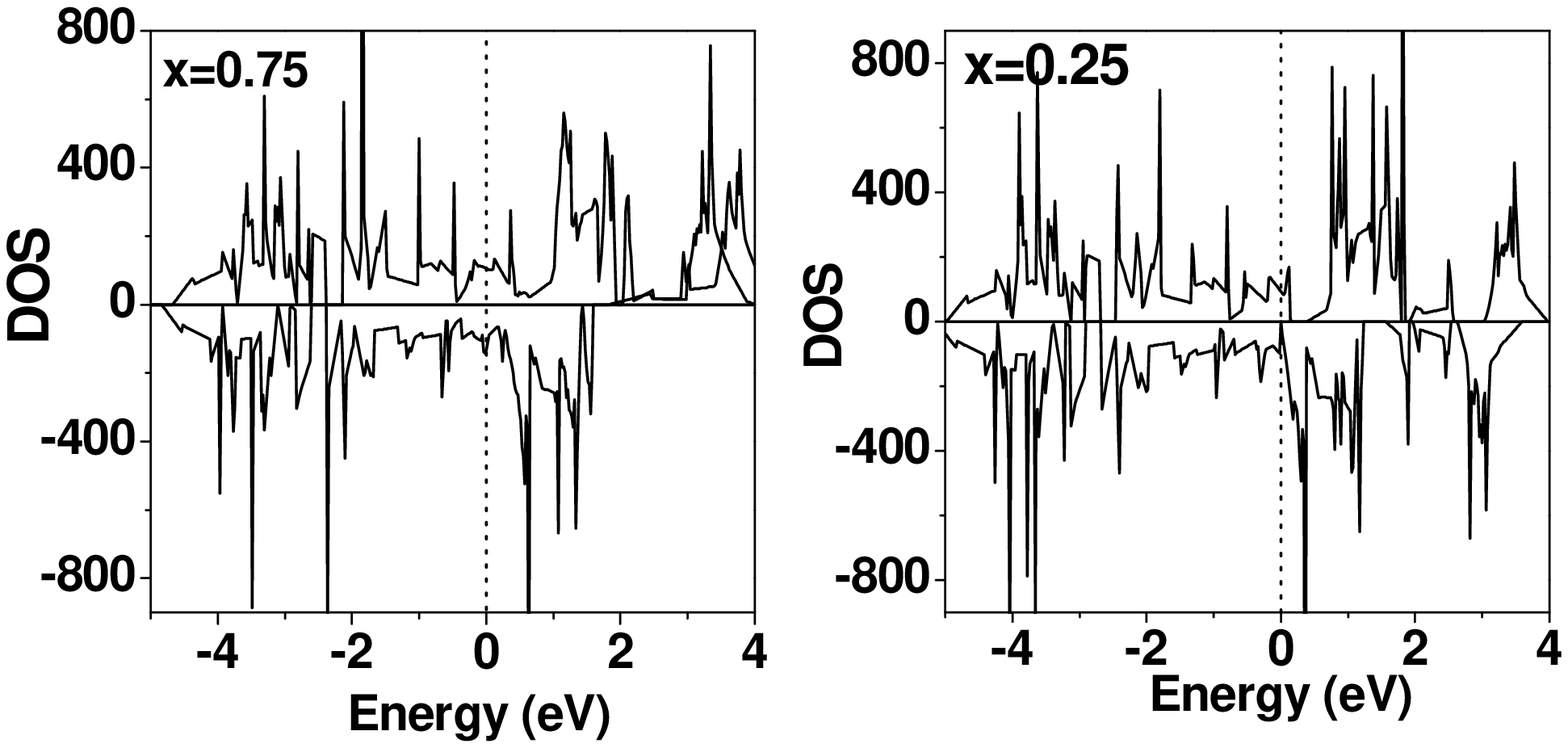}
\end{center}
\caption{The density of states for Na$_{0.75}$CoO$_{2}$ and Na$_{0.25}$CoO$%
_{2}$ with the enlarged separation of the CoO$_{2}$ layers $c=3.47$\AA. The
density of states is in atomic unit (Ryd)$^{-1}$. The up-spin DOS is shown
above the horizontal line and the down-spin below.}
\end{figure}

In conclusion, pseudopotential method with local spin density functional
approximation is used to calculate the electronic band structures of Na$_{x}$%
CoO$_{2}$ ($x=0.25$, $0.5$, $0.75$, and $1$) with the structure relaxations.
By increasing the Na content, both the hybridization between cobalt and
oxygen orbitals and the itinerant ferromagnetism are decreased, and a phase
transition is predicted from a wide-band ferromagnetic to narrow band
paramagnetic metals. It is the local distortions of the oxygen ions that
reduces the itinerant ferromagnetism as increasing the sodium content. We
also found that straining the CoO$_{2}$ layers enhances both the
hybridization and ferromagnetism. Whether the superconductivity of Na$%
_{0.35} $CoO$_{2}\cdot $1.3H$_{2}$O is related to a triplet (p-wave)
electron pairing is required further investigation.

This research was supported by National Key Program of Basic Research
Development of China (Grant No.G2000067107) and the National Natural Science
Foundation of China under Grant No.10274036.

\end{document}